# How predictable are symptoms in psychopathological networks? A reanalysis of 18 published datasets

**J. M. B. Haslbeck\* and E. I. Fried**

*Department of Psychology, University of Amsterdam, The Netherlands*

**Background.** Network analyses on psychopathological data focus on the network structure and its derivatives such as node centrality. One conclusion one can draw from centrality measures is that the node with the highest centrality is likely to be the node that is determined most by its neighboring nodes. However, centrality is a *relative* measure: knowing that a node is highly central gives no information about the extent to which it is determined by its neighbors. Here we provide an *absolute* measure of determination (or controllability) of a node – its predictability. We introduce predictability, estimate the predictability of all nodes in 18 prior empirical network papers on psychopathology, and statistically relate it to centrality.

**Methods.** We carried out a literature review and collected 25 datasets from 18 published papers in the field (several mood and anxiety disorders, substance abuse, psychosis, autism, and transdiagnostic data). We fit state-of-the-art network models to all datasets, and computed the predictability of all nodes.

**Results.** Predictability was unrelated to sample size, moderately high in most symptom networks, and differed considerable both within and between datasets. Predictability was higher in community than clinical samples, highest for mood and anxiety disorders, and lowest for psychosis.

**Conclusions.** Predictability is an important additional characterization of symptom networks because it gives an absolute measure of the *controllability* of each node. It allows conclusions about how self-determined a symptom network is, and may help to inform intervention strategies. Limitations of predictability along with future directions are discussed.

## Introduction

In the network approach to psychopathology, mental disorders are understood as networks of interacting symptoms, and by studying the structure of these networks one hopes to find explanatory models for the etiology of disorders and effective interventions (Cramer *et al.* 2010; Borsboom & Cramer, 2013). This perspective has provided new and intuitively appealing explanations of psychopathological phenomena, and has been applied to many different mental disorders (for a review, see Fried *et al.* 2016a).

While the analysis of the *structure* of symptom networks has led to important insights, in this paper we focus on another important characteristic that has not been considered so far in the literature: *predictability*, i.e. the degree to which a given node can be predicted by all other nodes in the network. Predictability is an important measure because it tells us on an interpretable absolute scale (e.g. 40% variance explained) how much a node is determined by other nodes in the network. Thereby, predictability gives us an idea of how clinically relevant connections (also called 'edges') are: if node A is connected to many other nodes, but these together explain only 1% of its variance, then these edges are not interesting in many situations. As an example, take the problem of selecting an intervention on insomnia in two hypothetical symptom networks in Fig. 1: in the network of the first patient (a), 80% of the variance of insomnia is explained by the two nodes that are connected to it, worrying and concentration problems, as indicated by the grey area in the ring around the node; it is plausible that an intervention on worrying may have a considerable impact on the sleep problems. In contrast, in the network of the second patient (b), insomnia is only weakly determined by its neighbors (11% variance explained), and an efficient intervention on insomnia via worrying seems questionable. Instead, we should search for relevant variables outside the current network that have an effect on insomnia, or may want to consider intervening directly on insomnia, e.g. by administering

\* Address for correspondence: J. M. B. Haslbeck, University of Amsterdam, Psychological Methods, Nieuwe Achtergracht 129-B, 1018 WT, Amsterdam, The Netherlands.

(Email: jonashaslbeck@gmail.com, http://jmbh.github.io)

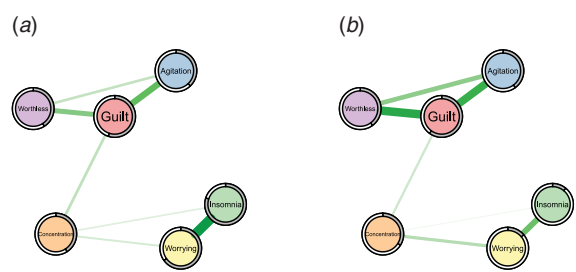

**Fig. 1.** Two example symptom networks with different predictability measures. Left: insomnia is strongly determined by the nodes connected to it (80% variance explained as indicated by the grey pie chart). Right: insomnia is weakly determined by the nodes connected to it (11% variance explained).

sleeping pills. Predictability thus helps us to estimate the potential success of clinical interventions on a symptom via the symptom network and could thereby guide treatment selection.

Clearly, predictability depends on the number and the strength of the edges a node is connected to: a node with many strong edges tends to have a higher predictability than a node with few weak edges. For instance, we can expect from the edge weights in Fig. 1 that insomnia is *better* predicted in network (a) than (b). However, we do not know *how well* we can predict insomnia on an absolute scale in either case. In contrast, predictability does provide such an absolute scale and thereby goes beyond the network structure and centrality indices reported in prior literature.

In summary, this work makes the following contributions:

1. We introduce predictability as an additional measure to characterize network models, and discuss the relationship between predictability and derivatives of the network structure, such as centrality measures.
2. We provide a reproducible example (the R-code and data can be found in the Online Supplementary Material) of how to estimate and interpret predictability in psychopathological networks using the data on bereavement and depression from Fried *et al.* (2015), serving as a tutorial for researchers.
3. We provide a first glimpse into predictability in the field of psychopathology by re-analyzing 25 datasets from 18 published papers that used network analyses. We discuss theoretical implications of the variability of predictability within and between networks and the relation between predictability and the network structure. In addition, we make all datasets we are allowed to share (5/25), as well as the weighted adjacency matrices (i.e. the network structures) of *all* 25 datasets available in the Online Supplementary material.

## Methods

### Literature review and data

We aimed to identify all papers in the field of psychopathology that applied network analysis techniques to cross-sectional data. To this end, we combined all papers known to the authors with the results of a literature review: we searched the databases PsycNET, ISI Web of Science and GoogleScholar using the names of the most prevalent mental disorders in combination with 'Network' as keywords. This literature review yielded 23 papers published between 2010 and 2016. We excluded one paper as the used data was identical to the data used in another paper. We contacted the authors of the remaining 22 papers and were able to obtain the data of the 18 papers described in Table 1. For further details about the literature review see Appendix A.1. The authors in the respective papers estimated Gaussian Graphical Models (GGMs) using the *qgraph* package (Epskamp *et al.* 2012), Ising models using the *IsingFit* package (van Borkulo *et al.* 2014), and the parameters for relative importance networks using the *relaimpo* package (Grömping, 2006, 2007) (see column 'original analysis' in Table 1). Datasets predominantly feature symptoms or clinical problems as nodes, although some contain contextual variables (e.g. *age of diagnosis* in Deserno *et al.* 2016).

### Statistical methods

We fitted GGMs to the continuous datasets and Ising models to the binary datasets. These models are considered the state-of-the-art and were also used in most of the papers included in our re-analysis (see Table 1). For an accessible tutorial on how to estimate GGMs, see Epskamp & Fried (2016). We computed predictability measures using the R-package *mgm* (Haslbeck & Waldorp, 2015; Haslbeck & Waldorp, 2016a). Note that in the case of GGMs, our estimation procedure was slightly different than the one in the original analyses as we did not estimate polychoric correlations before using the correlation matrix to estimate the graph structure using the graphical lasso (e.g. Epskamp *et al.* 2012). We instead used the neighborhood regression approach implemented in the *mgm* package, which is necessary to compute predictability. In the case of the Ising model, there are no differences since the node-wise estimation of *mgm* is identical to the node-wise estimation in *IsingFit* (van Borkulo *et al.* 2014). Note that the reported sample size in Table 1 in some cases differs from the one reported

**Table 1.** *Characteristics of papers included in the data reanalysis*

| Paper | Subfield | Datatype | $p$ | $N$ | Original analysis |
|---|---|---|---|---|---|
| Anderson *et al.* (2015) | Autism | Continuous | 14 | 477 | Correlation |
| Armour *et al.* (2017) | PTSD | Continuous | 27 | 221 | GGM |
| Beard *et al.* (2016) | Anxiety, Depression | Continuous | 17 | 1029 | GGM |
| Borsboom & Cramer (2013) | Anxiety, Depression | Binary | 18 | 9282 | Ising Model |
| Boschloo *et al.* (2016a) | General | Binary | 12 | 501 | Ising Model |
| Deserno *et al.* (2016) | Autism | Continuous | 27 | 301 | GGM |
| Fried *et al.* (2015) | Bereavement | Binary | 12 | 515 | Ising Model |
| Fried *et al.* (2016a, c) | Depression | Continuous | 28 | 3463 | GGM |
| Goekoop & Goekoop (2014) | General | Continuous | 63 | 192 | Correlation |
| Hoorelbeke *et al.* (2016) | Depression | Continuous | 6 | 69 | GGM |
| Koenders *et al.* (2015) | Bipolar | Continuous | 16 | 126 | Correlation |
| McNally *et al.* (2015) | PTSD | Continuous | 17 | 362 | GGM, relative importance |
| Rhemtulla *et al.* (2016) | Substance Abuse | Binary | 11 | 2405 | Ising model |
| Robinaugh *et al.* (2014) | Bereavement | Continuous | 19 | 1532 | GGM |
| Robinaugh *et al.* (2016) | Complicated Grief | Continuous | 13 | 195 | GGM |
| Ruzzano *et al.* (2015) | Autism | Binary | 17 | 213 | Correlation |
| Santos *et al.* (2017) | Depression | Continuous | 20 | 503 | GGM |
| Wigman *et al.* (2016) | Psychosis | Binary | 56 | 283 | GGM |

Notes: GGM, Gaussian Graphical Model. Datatype refers to the variables after preprocessing as performed in the original papers; this means that some datasets were actually on an ordinal scale with more than two categories, but were binarized for the analysis by the original papers (we did the same).

in the original study. In these cases, the authors deleted missing values pairwise to compute the sample covariance matrix and reported the full sample size. With the neighborhood regression approach, however, we have to delete missing values casewise, resulting in a smaller number of observations.

As predictability measures, we selected the *proportion of explained variance* for (centered) continuous variables and a *normalized accuracy measure* for binary variables. The normalized accuracy measure quantifies how a node is determined by its neighboring nodes *beyond* the intercept model. This is important, because for instance if a binary variable with 100 cases has 5 zeros and 95 ones, then the intercept model (which predicts a one for each case) alone would already lead to an accuracy of 95% without considering any other nodes. The normalized predictability measure takes this into account and is zero when other variables do not predict the node at hand beyond the intercept model; a more detailed explanation of both proportion of explained variance and normalized accuracy can be found elsewhere (Haslbeck & Waldorp, 2016b). Both measures range from 0 to 1: 0 means that we cannot at all predict a node by other nodes in the network, whereas 1 implies perfect prediction. In addition to predictability, we computed the following node centrality measures: weighted degree centrality, betweenness centrality, closeness centrality and eigenvector centrality (Newman *et al.* 2011).

Rhemtulla *et al.* (2016) split their data in six subgroups (abuse of cannabis, sedatives, stimulants, cocaine, opioids or hallucinogens) and Koenders *et al.* (2015) used three subgroups (mildly depressed, predominantly depressed, cycling). We followed the analyses in their papers and estimated six and three separate networks (see also Fig. 3), respectively. Overall, this led to 25 datasets/networks from 18 papers.

## Results

### Application example: node-wise predictability in data of Fried et al. (2015)

Before discussing the results of the re-analysis of all papers, we provide an example of how to estimate and interpret predictability using the depression and bereavement dataset analyzed in Fried *et al.* (2015); see Fig. 2. The Online Supplementary material contains R-code and data to reproduce Fig. 2.

Fried et al. (2015) re-analyzed the Changing Lives of Older Couples study (Carr *et al.* 2005). The network in Fig. 2 represents the cross-sectional network structure of 10 dichotomous depression symptoms (measured via the 10-item CES-D) and 1 condition node (*loss*), which codes whether participants belong to the bereaved group who had lost their spouse prior to this follow-up time point, or the still-married control group. Several results of the predictability analysis are noteworthy.

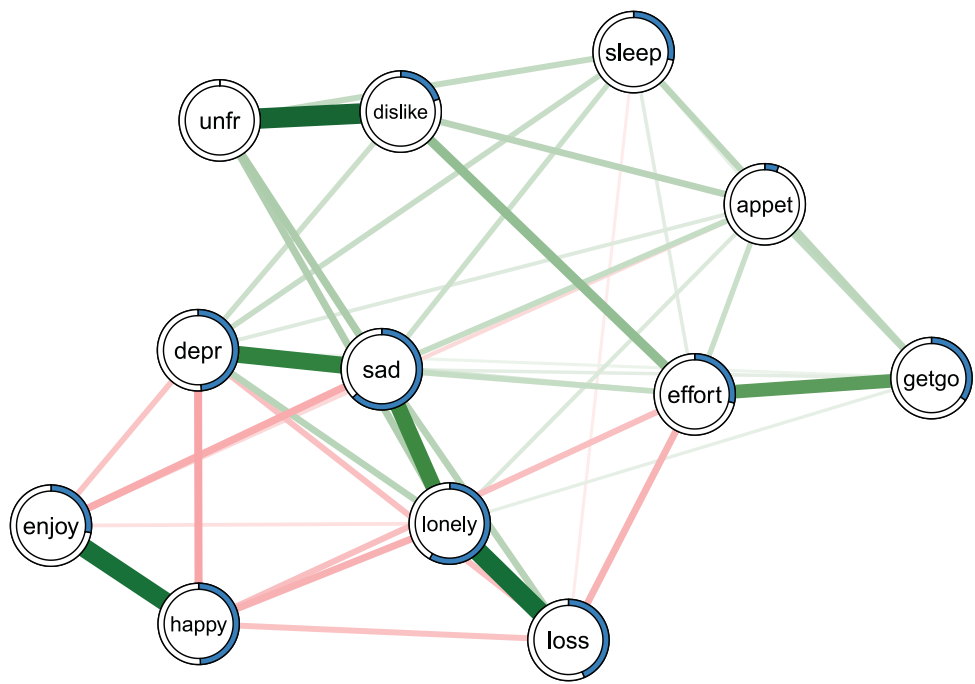

**Fig. 2.** Ising model estimated on the data of Fried *et al.* (2015). Green edges indicate positive relationships, red edges indicate negative relationships. The blue ring around each node represents its predictability. loss, spousal loss; depr, depressed; effort, everything is an effort; sleep, restless sleep; unfr, people are unfriendly; enjoy, enjoy life; appet, poor appetite; dislike, people dislike me; getgo, cannot get going. For a color version of this figure, see the digital version of the paper.

First, the average predictability across all nodes is 0.34, indicating that 34% of the variance of a node that is not predicted by the intercept model is explained by its neighbors. Compared with the predictability results of all other datasets (see below), this is an average level of predictability.

Second, *appet* (poor appetite) and *unfr* (people are unfriendly) stand out with the lowest predictability estimates in the network (0.06 and 0), implying that all other nodes together share nearly no variance with these items. The low predictability of poor appetite is consistent with psychometric studies of depression scales, showing that weight and appetite items often form a distinct cluster of nodes (or factor) and show only weak partial correlations with other depression symptoms (e.g. Fried *et al.* 2016*a, c*). In contrast, the low predictability *unfr* is likely explained by the low variance in this variable: 94% of the cases report the symptom to be absent. This leads to a situation in which the model including the neighbors gives the same predictions as the intercept model. Because the normalized predictability measure used here captures the predictability *beyond* the intercept model, we get a measure of zero (for details see Haslbeck & Waldorp, 2016*b*).

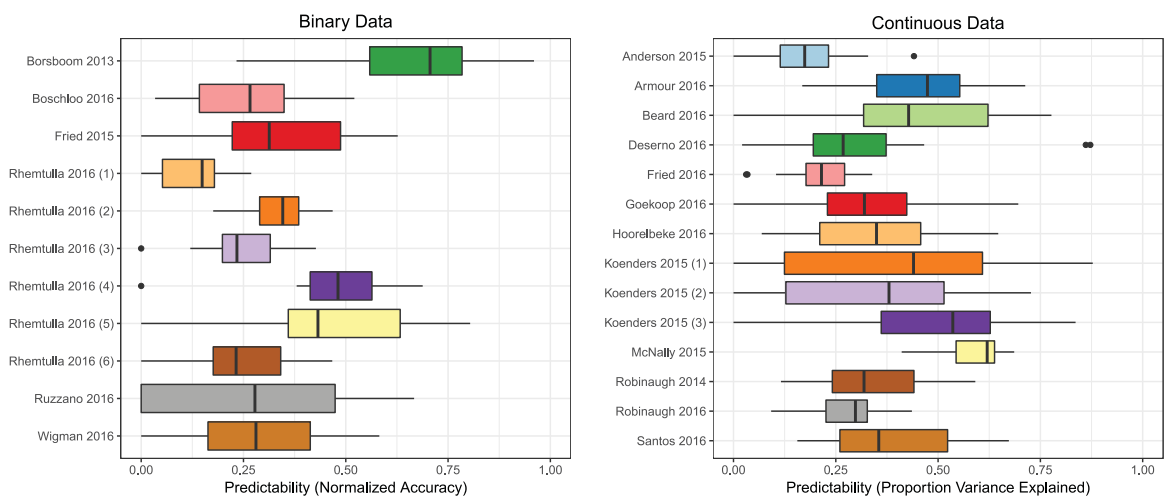

**Fig. 3.** Summaries of the distribution of predictability measures for datasets with continuous (left) and binary (right) data. The *x*-axis shows the predictability measure (ranging from 0 to 1): 'normalized accuracy' for binary variables and 'proportion of explained variance' for continuous variables. The box plot whiskers show 1.5 times the Interquantile Range (IQR). For a color version of this figure, see the digital version of the paper.

Finally, negative emotions such as *depr, sad* and *lonely* have comparably high predictability values (0.48, 0.63 and 0.58). This could either be due to the fact that these items measure different concepts that strongly influence each other, or because they capture similar constructs (e.g. *depr* and *sad* may tap into the same emotion).

### Re-analysis of 25 datasets

We now turn to the re-analysis of 25 datasets from 18 published papers in the psychopathology network literature. Figure 3 shows box plots describing the distributions of predictability measures for all included datasets. In general, we see that symptoms in networks can often be predicted reasonably well by all other symptoms in the network.

A few things stand out. First, node-predictability varies considerably *within* datasets, as can be seen by the spread in the distributions of predictability measures that is summarized in the boxplots: the bold vertical bar corresponds to the median, the box indicates the 25 and 75% quantiles, and the whiskers show 1.5 times the interquartile range.

Second, there is a considerable amount of node-predictability variation *across* datasets. This difference is not trivially explained by differences in sample size between datasets: the Spearman correlation between mean predictability and sample size is −0.27. In addition, we explored whether predictability differences across samples were related to severity of psychopathology. To that end, we classified all datasets into an ordinal variable indicating severity (0 = all healthy, 1 = mixed, 2 = clinical populations). The weighted (by number of observations) Spearman correlation between this severity variable and predictability was −0.81, providing evidence that networks of clinical samples may have a lower mean predictability than networks of healthy samples. This is consistent with findings of lower dimensionality of symptom networks of healthier patients (Fried *et al.* 2016*a*). However, these results are at best preliminary because a clear classification of datasets into predominantly healthy, sick, or mixed was difficult, and because we analyzed a small number of often highly heterogeneous datasets. Like many other results presented in this paper, these analyses serve as example of what research questions can be explored with predictability results, rather than strong evidence.

Third, it stands out that the six substance abuse subsamples of Rhemtulla *et al.* (2016) differ considerably in their mean predictability. A possible explanation for these differences is that the symptoms are consequences of a common cause – the consumed substance – and that the influence of this common cause is differentially strong for different substances (e.g. stronger for opioids than cannabis). A similar argument could be made for the datasets on posttraumatic stress disorder (PTSD): symptoms may co-vary (and hence predict each other well) because they are all caused by the traumatic experience. This contrasts with the network approach to psychopathology, and we will turn to this issue of (unobserved) common causes in the discussion.

Fourth, we observed a very high mean predictability for the depression network of Borsboom & Cramer (2013). We suspect that this pattern may come from skip questions that the authors replaced with 0s (i.e. no information on symptoms was coded as symptoms being absent). This procedure leads to spurious relationships, because variables become related via their shared missing value pattern that is determined by the structure of the skip questions (0s are imputed for multiple variables at the same time, inducing correlations among these items). We also observed a very high predictability of two items in the paper on autism by Deserno *et al.* (2016) (see the outliers in Fig. 3) – *age* and *age of diagnosis*. These have to be strongly correlated, because the former is an upper bound for the latter, i.e. a person cannot get a diagnosis at the age of 15 if the person is 9 years old.

Finally, the question arises whether predictability differs consistently across different types of datasets, for instance, across mental disorders. Differences in predictability across mental disorders can be interpreted as evidence for how self-determined a symptom network is: if predictability is high, the symptoms are largely determined by each other, if predictability is low, symptoms are largely influenced by additional variables (e.g. biological, environmental or additional symptoms) that are not included in the network. Figure 3 suggests that symptoms of depression and anxiety disorders might be more self-determined (average predictability = 0.42), while the symptoms of psychosis might be determined to a larger degree by other influences such as genes or environmental variables (0.28). Other explanations for the pattern of findings could be that the measurement error is larger for symptoms of psychosis, or that depression and anxiety assess very similar problems multiple times, which increases their respective predictability. Apart from comparing predictability across types of mental disorders, we could also investigate whether the predictability is higher for female *v.* male, or younger *vs.* older patients. While we do not have sufficient data to answer these questions, Fig. 3 provides numerous possibilities that should be investigated in more focused future studies.

### Relationship between predictability and edge weights

It is clear that there has to be a close relationship between the predictability of a node and the edge weights connected to that node: if a node is unconnected, its predictability by other nodes has to be zero. And the more edges are connected to a node, the higher its predictability *tends* to be. We illustrate this relationship using weighted degree centrality (the sum of absolute edge-weights connected to a node), which had the highest mean correlation with predictability (0.74, 0.67, 0.10 and 0.01 for weighted degree centrality, eigenvalue centrality, closeness centrality and betweenness centrality, respectively): in Fig. 4, we plot weighted degree against predictability, for all datasets shown in Fig. 3.

Each point corresponds to one node and its color indicates to which dataset it belongs (see Fig. 3). As expected, we observe a positive relationship between the centrality of a node and its predictability. This relationship is stronger for continuous-Gaussian variables, because here the edge weights (which are partial correlations in this special case) are always between −1 and 1, whereas edge weights in the Ising model for binary data are only constrained to be finite. However, the relationship is far from perfect: for example, for continuous-Gaussian variables, a centrality measure of 0.5 can coincide with a predictability measure between 0.1 and 0.5 and for binary variables, a centrality measure of 4 can coincide with a predictability measure between 0.1 and 0.8.

It is crucial to note, however, that centrality gives us only relative information about predictability: even if both measures would be correlated perfectly, we could only *order* all nodes by predictability, but we would not know the *absolute value* of the predictability of any node. This is similar to the correlation of the actual height of children in a classroom with their position in an ordering by height: these two metrics may be considerably related, but we can never know how tall Alice is from knowing she is the fifth tallest girl in class.

It would be possible to fit a regression model to predict predictability from degree centrality. However, both the mean predictability (see Fig. 3) and the strength of the linear relationship between both measures (see Fig. 4) differ greatly between datasets, which implies that a predictability inferred from centrality would be highly inaccurate. Given that predictability can be easily computed with freely available software (see Haslbeck & Waldorp, 2016*a*), we see no reason to accept these inaccuracies.

### Discussion

We showed that predictability is an important characteristic of network models in addition to their structure. Furthermore, we provided a first overview of how high predictability typically is in the field of psychopathology and suggest that analyzing predictability across disorders and groups of individuals may generate new theoretical insights.

Predictability was moderately high in most datasets, indicating that a considerable amount of the variation of nodes can be explained by other nodes in the network. We found that the average predictability was higher for certain disorders (e.g. depression, anxiety and PTSD) than for others (e.g. psychosis). This suggests that the symptom network of the former disorders is more self-determined, while nodes for the latter disorders are more strongly influenced by other factors that are not included in the network, such as additional symptoms or biological and environmental variables. We thus see predictability as a first attempt towards characterizing the *controllability* of the symptom network: if predictability is high, we can control symptoms via their neighboring symptom in the network – if it is low, we have to search for additional variables or intervene on the symptom directly. If our findings of low predictability for specific disorders or groups of patients can be replicated in future studies, this calls for research on important variables beyond common symptoms.

In clinical practice, predictability enables us to judge the efficacy of a planned treatment: if the neighbors of symptom A explain 90% of its variance, an intervention on symptom A via its neighbors seems viable. In contrast, if they explain only 5% of the variance, one would rather search for additional variables outside the network or try to intervene on the node directly (instead of trying to control the node via neighboring nodes).

It is important to note several limitations of the present paper. First, we only analyzed a small and heterogeneous sample of datasets (all available datasets we could obtain for this project), and a much larger database of studies is required to draw any strong conclusions when comparing, for instance, the predictability of different types of mental disorders. Due to the increasing popularity of network models in psychopathology, we look forward to having more data available in the next few years to tackle these and related questions.

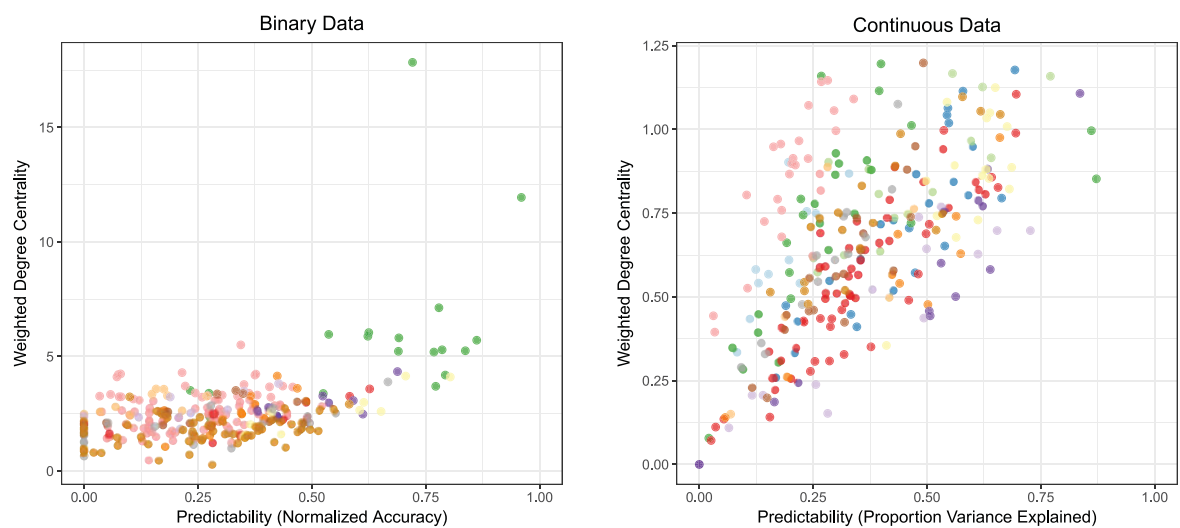

**Fig. 4.** The relationship between weighted degree centrality (*x*-axis) and predictability (*y*-axis) of each node in the datasets with continuous (left) and binary (right) data. The colors of the points correspond to the colors used for different papers in Fig. 3. For a color version of this figure, see the digital version of the paper.

Second, the present paper explored how well node A is predicted by all its neighbors. Another interesting question is how well node A is predicted by one particular neighboring node B. Unfortunately, there is no straightforward solution to this problem in the case of correlated predictors, which is nearly always the case in psychopathology data. For continuous-Gaussian data, solutions to this problem have been proposed that are based on variance decomposition (Grömping, 2007), and there are more general methods in the machine learning literature based on replacing a predictor by noise and investigating the drop in predictability (e.g. Breiman, 2001). While the performance of these methods is not always clear and requires further work, it would be important to extend and apply these approaches to the network models used in psychopathology research. From this limitation follows that we cannot quantify the 'predictive power' or 'degree of determination' of a given node on its neighboring nodes on an absolute scale (the causal opposite of predictability or controllability). However, if predictability is low for all nodes in the network, we do know that no node exerts a strong influence on any other node in the network.

Third, when calculating predictability of node A, we assume that all edges are directed towards that node A, i.e. that all adjacent nodes are causing A, but not vice versa. However, we do not know whether this is true because the direction of edges is generally unknown in cross-sectional data (Pearl, 1991). It follows that the predictability of a node is an *upper bound* for how much it is determined by the nodes it is connected to. While it is important to keep this limitation in mind, it may not matter that much in many situations: for instance, if the predictability of symptom A is too low to render an intervention via neighboring symptoms viable, it does not matter that the true predictability is probably lower. The predictability estimate can be improved by any method that reliably replaces our assumption about the direction of edges by estimates about the direction. In cross-sectional data, the direction of edges can under a set of stringent assumptions be inferred via causal search algorithms such as the PC-algorithm (Spirtes *et al.* 2000). In time-series data with lagged effects, this problem is circumvented by using the direction of time: if A and B are related and A precedes B in time, then we say that A Granger-causes B and we have a directed edge from A to B (Granger, 1969). The predictability measure we propose here can easily be applied to these time-series models as well (see Haslbeck & Waldorp, 2016b).

Fourth, the interpretation of predictability of a node as the degree to which it is determined by the node it is connected to is only appropriate if the network model is an appropriate model for the phenomenon at hand. A network model can be problematic or even inappropriate for a number of reasons (see also Haslbeck & Waldorp, 2016a; Fried & Cramer, 2016b; Fried *et al.* 2017). In the presence of two or more variables that measure the same underlying construct (e.g. several questions about sad mood) we would not interpret connections between those variables as genuine causal relations and hence we also would not interpret predictability as a measure of determination. Another problem is a situation in which variables are deterministically related such as the variables *age* and *age of diagnosis* in the paper of Deserno *et al.* (2016). Clearly in this case, we would not think of a process in which *age* is causing *age of diagnosis* or vice versa. Moreover, it is problematic if two or more nodes have a common cause that is not included in the network, because this leads to a spurious edge between these nodes. In all three cases, interpreting edges as genuine cause-effect relationships is incorrect, and the interpretation of predictability as degree of determination by neighboring nodes does not apply. This could be the case for substance abuse and PTSD where substance use and traumatic experiences may be common causes for (parts of) the symptom network (Fried & Cramer, 2016b). While this is a major limitation, it applies to any other statistical model as well: for instance, interpreting Cronbach's alpha or factor loadings in factor models makes only sense in case the factor model is the appropriate model for the data.

In sum, predictability is a useful additional characterization of psychopathological networks, may have direct implications for clinical practice, and provides a method to investigate theoretical questions such as the degree of self-determination of a network.

## Supplementary material

The supplementary material for this article can be found at https://doi.org/10.1017/S0033291717001258.

## Acknowledgements

The authors would like to thank Kamran Afzali, Denny Borsboom, Gary Brown, Tiago Cabaço, Sara Plakolm, Mijke Rhemtulla and Lourens Waldorp for helpful comments. This project was supported by the European Research Council Consolidator Grant no. 647209.

## Declaration of interest

None.

## Appendix A 1: detailed description of literature review

We performed a literature search on the databases PsycNET, ISI Web of Science and GoogleScholar using 'Network AND X' as a keyword, where we made nine separate searches, where X was either 'Psychopathology', 'Comorbidity', 'Post Traumatic Stress Disorder', 'Depression', 'Anxiety', 'Schizophrenia', 'Psychosis', 'Personality Disorder', or 'Substance'. We constrained our search to the period 2010–2016 as we consider the paper of Cramer *et al.* (2010) as the first 'network paper' in the field of psychopathology. While we checked all search results for PsycNET and ISI Web of Science, for GoogleScholar we only went through the first 10 pages of results, because going through all results was not feasible [e.g. the search query 'Network + Psychopathology' led to 187 000 results (10/7/2016)].

This literature review yielded 23 papers. We excluded one paper as the used data was identical to the data used in another paper. We contacted the first authors of the remaining 22 papers end of July. Authors who did not respond to our initial email were reminded end of August. We consider all datasets we received up until 15 November, giving the authors over 2 months to share their data. Within this 2-month period, we were able to obtain the datasets of 18/22 papers. Table A1 states the reason for not obtaining the data of four papers.

**Table A1.** *lists the 23 papers we found by combining papers the authors knew of with the additional papers found in the literature Review. The papers Cramer et al. (2010) and Borsboom & Cramer (2013) use the same dataset; therefore we only included Borsboom & Cramer (2013)*

| Paper | Result of data request |
| --- | --- |
| Anderson *et al.* (2015) | Obtained |
| Armour *et al.* (2017) | Obtained |
| Beard *et al.* (2016) | Obtained |
| Borsboom & Cramer (2013) | Obtained |
| Boschloo *et al.* (2015) | New policy of US National Institute on Alcohol Abuse and Alcoholism does not allow sharing data anymore (personal communication) |
| Boschloo *et al.* (2016b) | Obtained |
| Boschloo *et al.* (2016b) | Requirements to obtain data from NESDA for re-analysis unfeasible for this project |
| Curtiss & Klemanski (2016) | Did not share their data |
| Deserno *et al.* (2016) | Obtained |
| Fried *et al.* (2015) | Obtained |
| Fried *et al.* (2016a, c) | Obtained |
| Goekoop & Goekoop (2014) | Obtained |
| Hoorelbeke *et al.* (2016) | Obtained |
| Koenders *et al.* (2015) | Obtained |
| McNally *et al.* (2015) | Obtained |
| Rhemtulla *et al.* (2016) | Obtained |
| Robinaugh *et al.* (2014) | Obtained |
| Robinaugh *et al.* (2016) | Obtained |
| Ruzzano *et al.* (2015) | Obtained |
| Santos *et al.* (2017) | Obtained |
| van Borkulo *et al.* (2014) | Requirements to obtain data from NESDA for re-analysis unfeasible for this project |
| Wigman *et al.* (2016) | Obtained |